\documentclass[a4paper,10pt,twoside]{cpc-hepnp}

\usepackage{multicol}
\usepackage{graphicx}
\usepackage{booktabs}
\usepackage{amssymb,bm,mathrsfs,bbm,amscd}
\usepackage[tbtags]{amsmath}
\usepackage{lastpage}

\begin{document}


\footnotetext[0]{Received 31 December, 2009}

\title{AdS/QCD and Light Front Holography: A New Approximation to QCD\thanks{Supported by Department of Energy Department
of Energy  contract DE--AC02--76SF00515.  SLAC-PUB-13876 }}

\author{%
 ~Stanley J. Brodsky$^{1}$\email{sjbth@slac.stanford.edu}%
%
\quad Guy F. de T\'eramond$^{2}$\email{gdt@asterix.crnet.cr}%
}
\maketitle

\address{%
1~(SLAC National Accelerator Laboratory,
Stanford University, Stanford, CA 94309, USA)\\
2~(Universidad de Costa Rica, San Jos\'e, Costa Rica)
\\
}

\begin{abstract}
The combination of Anti-de Sitter  space (AdS) methods with light-front holography leads to a semi-classical first approximation to the spectrum and wavefunctions of meson and baryon light-quark  bound states.
Starting from the bound-state Hamiltonian equation of motion in QCD, we derive  relativistic light-front wave equations in terms of an invariant impact variable $\zeta$ which measures the separation of the quark and gluonic constituents within the hadron at equal light-front time. These equations of motion in physical space-time are  equivalent to the equations of motion which describe the propagation of spin-$J$ modes in anti--de Sitter (AdS) space. Its eigenvalues give the hadronic spectrum, and its eigenmodes represent the probability distributions of the hadronic constituents at a given scale. Applications to the light meson and baryon spectra are presented. The predicted  meson spectrum has a string-theory Regge form ${\cal M}^2 = 4 \kappa^2(n+L+S/2 )$; {\it i.e.}, the square of the eigenmass is linear in both $L$ and $n$, where $n$ counts the number of nodes  of the wavefunction in the radial variable $\zeta$. The space-like pion and nucleon form factors are also well reproduced.
One thus obtains a remarkable
connection between the description of hadronic modes in AdS space and
the Hamiltonian formulation of QCD in physical space-time quantized
on the light-front  at fixed light-front time $\tau.$  The model
can be systematically improved  by using its complete orthonormal solutions to diagonalize the full QCD light-front Hamiltonian or by applying the Lippmann-Schwinger method in order to systematically include the QCD interaction terms.
\end{abstract}



\begin{multicols}{2}

\section{Introduction}
A long-sought goal in hadron physics is to find a simple analytic first approximation to QCD analogous to the Schr\"odinger-Coulomb equation of atomic physics.  This problem is particularly challenging since the formalism must be relativistic, color-confining, and consistent with chiral symmetry.

We have recently shown that  the combination of Anti-de Sitter  space (AdS) methods with light-front (LF) holography leads to a remarkably accurate first approximation for the spectra and wavefunctions of meson and baryon light-quark  bound states. The resulting equation for the mesonic $q \bar q$ bound states at fixed light-front time $\tau= t +z/c$ , the time marked by the
front of a light wave,~\cite{Dirac:1949cp}  has the form of a relativistic Lorentz invariant  Schr\"odinger equation
\begin{equation} \label{eq:QCDLFWE}
\left(-\frac{d^2}{d\zeta^2}
- \frac{1 - 4L^2}{4\zeta^2} + U(\zeta) \right)
\phi(\zeta) = \mathcal{M}^2 \phi(\zeta),
\end{equation}
where the confining potential is $ U(\zeta) = \kappa^4 \zeta^2 + 2 \kappa^2(L+S-1)$
in a soft dilaton modified background.
There is only one parameter, the mass scale $\kappa \sim 1/2$ GeV, which enters the confinement potential. Here $S=0,1$ is the spin of the $q $ and $ \bar q $, $L$ is their relative orbital angular momentum as determined in the light-front formalism and $\zeta = \sqrt{x(1-x) {\bf b}^2_\perp}$ is a Lorentz invariant coordinate which measures
the distance between the quark and antiquark; it is analogous to the radial coordinate $r$ in the Schr\"odinger equation.   In effect $\zeta$ represents the off-light-front energy shell or invariant mass dependence of the bound state; it  allows the separation of the dynamics of quark and gluon binding from
the kinematics of constituent spin and internal orbital angular momentum.~\cite{deTeramond:2008ht}

We thus obtain  a  single-variable LF relativistic Schr\"odinger
equation  which determines the spectra and light-front wavefunctions
(LFWFs) of hadrons for general spin and orbital angular momentum. This
LF wave equation serves as a semiclassical first approximation to QCD,
and it is equivalent to the equations of motion which describe the
propagation of spin-$J$ modes in  AdS space.  Light-front holography
thus provides a remarkable connection between the description of
hadronic modes in AdS space and the Hamiltonian formulation of QCD in
physical space-time quantized on the light-front  at fixed LF
time $\tau.$

\section{Mesons in Light-Front AdS/QCD}

The meson spectrum predicted by  Eq. \ref{eq:QCDLFWE} has a string-theory Regge form
${\cal M}^2 = 4 \kappa^2(n+ L+S/2)$; {\it i.e.}, the square of the eigenmasses are linear in both $L$ and $n$, where $n$ counts the number of nodes  of the wavefunction in the radial variable $\zeta$.  This is illustrated for the pseudoscalar and vector meson spectra in Figs. \ref{pion} and \ref{VM},
where the data are from Ref.~\cite{Amsler:2008xx}
The pion ($S=0, n=0, L=0$) is massless for zero quark mass, consistent with the chiral invariance of massless QCD.  Thus one can compute the hadron spectrum by simply adding  $4 \kappa^2$ for a unit change in the radial quantum number, $4 \kappa^2$ for a change in one unit in the orbital quantum number  $L$ and $2 \kappa^2$ for a change of one unit of spin $S$. Remarkably, the same rule holds for three-quark baryons as we shall show below.

\begin{center}
\includegraphics[angle=0,width=8cm]{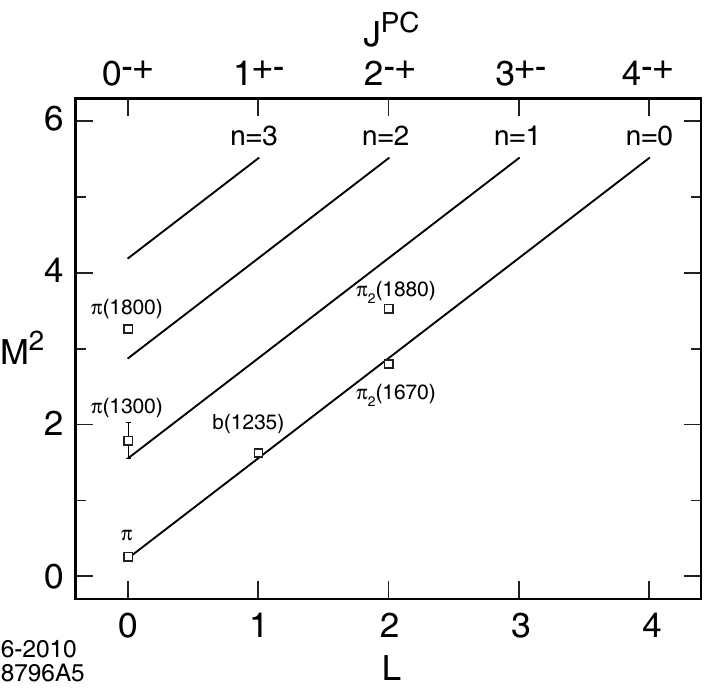}
\figcaption{\label{pion}Parent and daughter Regge trajectories for the $\pi$-meson family for
$\kappa= 0.6$ GeV.}
\end{center}

\begin{center}
\includegraphics[angle=0,width=8cm]{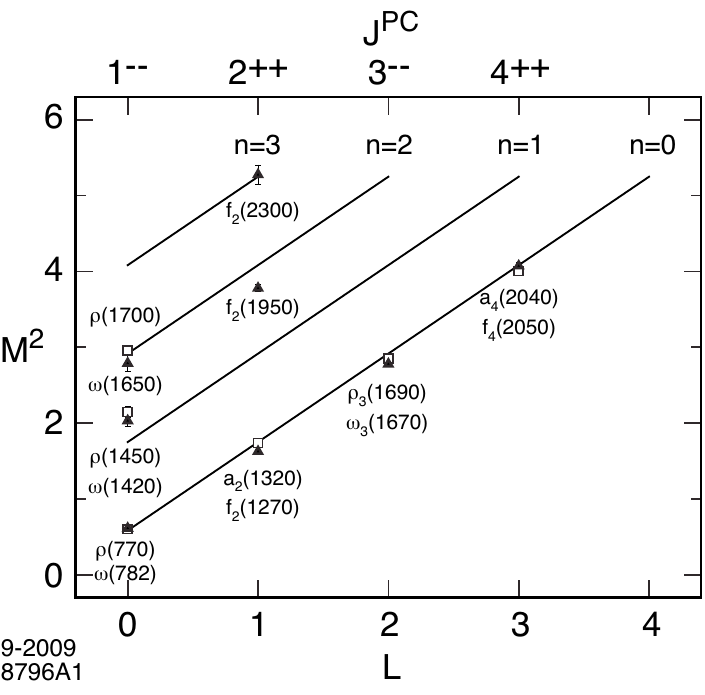}
\figcaption{\label{VM} Parent and daughter Regge Regge trajectories for the  $I\!=\!1$ $\rho$-meson
 and the $I\!=\!0$  $\omega$-meson families for $\kappa= 0.54$ GeV.}
\end{center}

The LFWFs of bound states in QCD are relativistic generalizations of the Schr\"odinger wavefunctions of atomic physics, but they are determined
at fixed light-cone time $\tau  = t +z/c$ -- the ``front form'' introduced by Dirac~\cite{Dirac:1949cp} -- rather than at fixed ordinary time
$t.$ It is natural to set boundary conditions at fixed $\tau$ and then evolve the system using the light-front (LF) Hamiltonian $P^-  \!= \!
P^0-P^3 = i {d/d \tau}$.  The invariant Hamiltonian $H_{LF} = P^+ P^- \! - P^2_\perp$ then has eigenvalues $\mathcal{M}^2$ where $\mathcal{M}$
is the physical mass.   Its eigenfunctions are the light-front eigenstates whose Fock state projections define the light-front wavefunctions.
The eigensolutions of  Eq. \ref{eq:QCDLFWE} provide the light-front wavefunctions of the valence Fock state of the hadrons $\psi(x,
\bf{b}_\perp)$  as illustrated for the pion in Fig. \ref{LFWF} for the soft and hard wall models.   The resulting distribution amplitude has a
broad form $\phi_\pi(x) \sim \sqrt{x(1-x)}$ which is compatible with moments determined from lattice gauge theory. One can then immediately
compute observables such as hadronic form factors (overlaps of LFWFs), structure functions (squares of LFWFS), as well as the generalized parton
distributions and distribution amplitudes which underly hard exclusive reactions. For example, hadronic form factors can be predicted from the
overlap of LFWFs, as in the Drell-Yan West formula. The prediction for the space-like pion form factor is shown in Fig.  \ref{PionFFSL}. The
pion form factor and the vector meson poles residing in the dressed current in the soft wall model require choosing  a value of $\kappa$ smaller
by a factor of $1/\sqrt 2$  than the canonical value of  $\kappa$ which determines the mass scale of the hadronic spectra.  This shift is
apparently due to the fact that the transverse current in $e^+ e^- \to q \bar q$ creates a quark pair with $L^z= \pm 1$ instead of the $L^z=0$
$q \bar q$ composition of the vector mesons in the spectrum.

\begin{center}
\includegraphics[width=6.8cm]{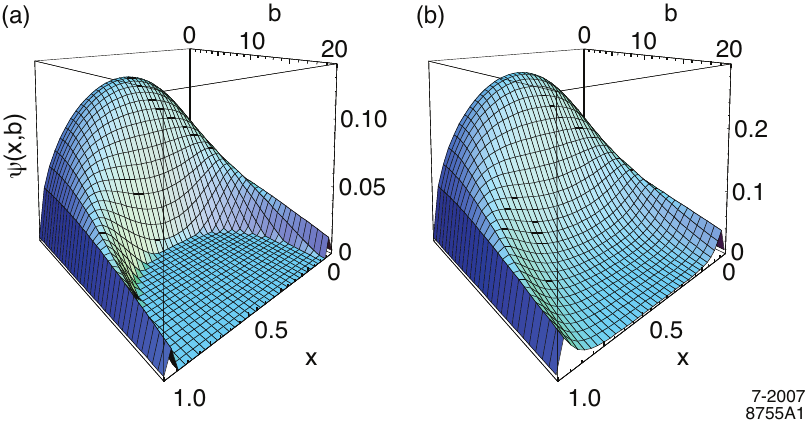}
 \figcaption{\label{LFWF} Pion LF wavefunction $\psi_\pi(x, \bf{b}_\perp$) for the  AdS/QCD (a) hard-wall ($\Lambda_{QCD} = 0.32$ GeV) and (b) soft-wall  ($\kappa = 0.375$ GeV)  models.}
\end{center}

\begin{center}
\includegraphics[width=6.8cm]{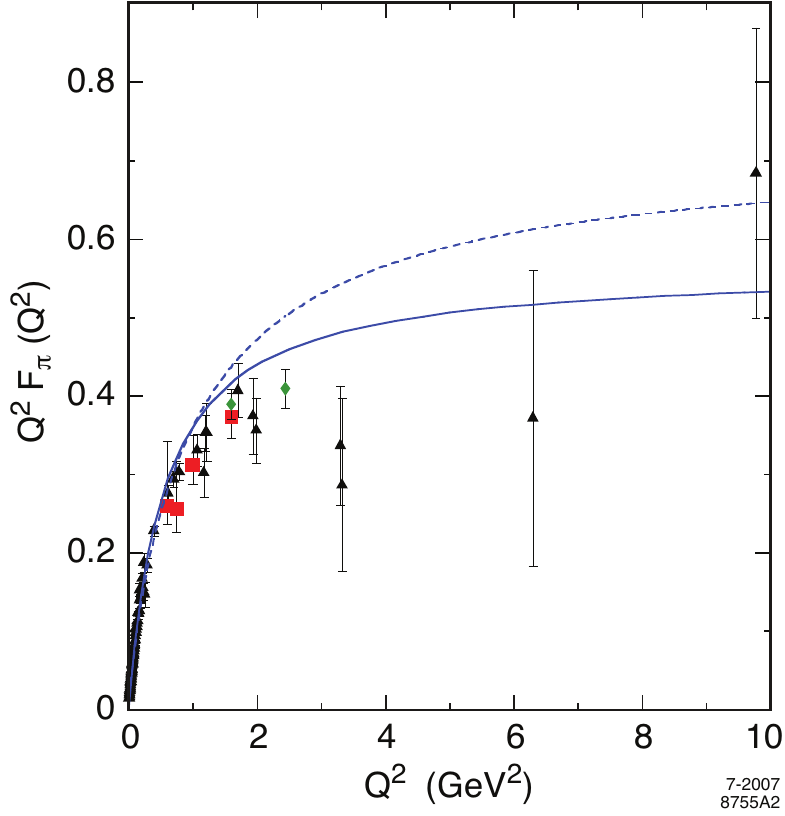}
\figcaption{ \label{PionFFSL} Space-like scaling behavior for $Q^2 F_\pi(Q^2).$ The continuous line is the prediction of the soft-wall model for
$\kappa = 0.375$ GeV.  The dashed line is the prediction of the hard-wall model for $\Lambda_{\rm QCD} = 0.22$ GeV. The triangles are the data
compilation of Baldini }
\end{center}

In the standard applications of  AdS/CFT methods, one begins with Maldacena's duality between  the conformal supersymmetric $SO(4,2)$
 gauge theory and a semiclassical supergravity string theory defined in a 10 dimension
 AdS$_5 \times S^5$
 space-time.~\cite{Maldacena:1997re} The essential mathematical tool underlying Maldacena's observation is the fact that the effects of scale transformations in a conformal theory can be mapped to the $z$ dependence of amplitudes in AdS$_5$ space.
QCD is not conformal but  there is in fact much empirical evidence from lattice, Dyson Schwinger theory and effective charges
that the QCD $\beta$ function vanishes in the infrared.~\cite{Deur:2008rf}  The QCD infrared fixed point arises since the propagators of the confined quarks and gluons in the  loop integrals contributing to the $\beta$ function have a maximal wavelength.~\cite{Brodsky:2008be} The decoupling of quantum loops in the infrared is analogous to QED where vacuum polarization corrections to the photon propagator decouple at $Q^2 \to 0$.

We thus begin with a conformal approximation to QCD to model an effective dual gravity description in AdS space. One uses the  five-dimensional AdS$_5$ geometrical representation of the conformal group to represent scale transformations within the conformal window. Confinement can be
effectively introduced with a sharp cut-off in the infrared region of AdS space, the ``hard-wall" model,~\cite{Polchinski:2001tt}
 or, more successfully,  using a dilaton background in the fifth dimension which produces a smooth cutoff and linear Regge trajectories,
the ``soft-wall" model.~\cite{Karch:2006pv}
The soft-wall AdS/CFT model with a positive-sign dilaton-modified AdS space leads to the
 potential $U(z) = \kappa^4 z^2 + 2 \kappa^2(L+S-1),$~\cite{deTeramond:2009xx}
 We assume a dilaton profile $\exp(+\kappa^2 z^2)$~\cite{deTeramond:2009xx, Andreev:2006ct, Zuo:2009dz}, with  opposite sign  to that of Ref.~\cite{Karch:2006pv}.

Glazek and Schaden~\cite{Glazek:1987ic} have shown that a  harmonic oscillator confining potential naturally arises as an effective potential between heavy quark states when one stochastically eliminates higher gluonic Fock states. Also, Hoyer~\cite{Hoyer:2009ep} has argued that the Coulomb  and  a linear  potentials are uniquely allowed in the Dirac equation at the classical level. The linear potential  becomes a harmonic oscillator potential in the corresponding Klein-Gordon equation.

Individual hadrons in AdS/QCD are identified by matching the power behavior of the hadronic amplitude at the AdS boundary at small $z$ to the twist of its interpolating operator at short distances $x^2 \to 0$, as required by the AdS/CFT dictionary. The twist corresponds to the dimension of fields appearing in chiral super-multiplets;~\cite{Craig:2009rk}
thus the twist of a hadron equals the number of constituents plus the relative orbital angular momentum.
We  then apply light-front holography to relate the amplitude eigensolutions  in the fifth dimension coordinate $z$  to the LF wavefunctions in the physical spacetime variable  $\zeta$.
Light-Front Holography can
be derived by observing the correspondence between matrix elements obtained in AdS/CFT~\cite{Polchinski:2002jw} with the corresponding formula using the LF
representation.~\cite{Brodsky:2006uqa, Brodsky:2007hb}   Identical results are obtained from the mapping of the QCD gravitational form factor
with the expression for the hadronic gravitational form factor in AdS space,~\cite{Brodsky:2008pf, Abidin:2008ku} a nontrivial test of consistency.

Equation (\ref{eq:QCDLFWE}) was derived by taking the LF bound-state Hamiltonian equation of motion as the starting
point.~\cite{deTeramond:2008ht} The term $L^2/ \zeta^2$  in the  LF equation of motion  (\ref{eq:QCDLFWE})
is derived from  the reduction of the LF kinetic energy when one transforms to the radial $\zeta$  and angular coordinate
$\varphi$, in analogy to the $\ell(\ell+1)/ r^2$ Casimir term in Schr\"odinger theory.  One thus establishes the interpretation of $L$ in the AdS equations of motion.
The interaction terms build confinement and correspond to
truncation of AdS space~\cite{deTeramond:2008ht} in an effective dual gravity  approximation.
The duality between these two methods provides a direct
connection between the description of hadronic modes in AdS space and
the Hamiltonian formulation of QCD in physical space-time quantized
on the light-front  at fixed LF time $\tau.$

The identification of orbital angular momentum of the constituents is a key element in the description of the internal structure of hadrons using holographic principles. In our approach  quark and gluon degrees of freedom are explicitly introduced in the gauge/gravity correspondence,~\cite{Brodsky:2003px} in contrast with the usual
AdS/QCD framework~\cite{Erlich:2005qh,DaRold:2005zs} where axial and vector currents become the primary entities as in effective chiral theory.

Unlike the top-down string theory approach,  one is not limited to hadrons of maximum spin
$J \le 2$, and one can study baryons with finite color $N_C=3.$   Higher spin modes follow from shifting dimensions in the AdS wave equations.
In the soft-wall
model the usual Regge behavior is found $\mathcal{M}^2 \sim n +
L$, predicting the same multiplicity of states for mesons
and baryons as observed experimentally.~\cite{Klempt:2007cp}
It is possible to extend the model to hadrons with heavy quark constituents
by introducing nonzero quark masses and short-range Coulomb
corrections.  For other
recent calculations of the hadronic spectrum based on AdS/QCD, see Refs.~\cite{Boschi-Filho:2002vd,   BoschiFilho:2005yh, Evans:2006ea, Hong:2006ta, Colangelo:2007pt, Forkel:2007ru, Vega:2008af, Nawa:2008xr, dePaula:2008fp,  Colangelo:2008us, Forkel:2008un, Ahn:2009px, Sui:2009xe}. Other recent computations of the pion form factor are given in
Refs.~\cite{Kwee:2007dd, Grigoryan:2007wn}.

\section{Baryons in Light-Front AdS/QCD}

For baryons, the light-front wave equation is a linear equation
determined by the LF transformation properties of spin 1/2 states. A linear confining potential
$U(\zeta) \sim \kappa^2 \zeta$ in the LF Dirac
equation leads to linear Regge trajectories.~\cite{Brodsky:2008pg}   For fermionic modes the light-front matrix
Hamiltonian eigenvalue equation $D_{LF} \vert \psi \rangle = \mathcal{M} \vert \psi \rangle$, $H_{LF} = D_{LF}^2$,
in a $2 \times 2$ spinor  component
representation is equivalent to the system of coupled linear equations
\begin{eqnarray} \label{eq:LFDirac} \nonumber
- \frac{d}{d\zeta} \psi_- -\frac{\nu+{1\over 2}}{\zeta}\psi_-
- \kappa^2 \zeta \psi_-&=&
\mathcal{M} \psi_+, \\ \label{eq:cD2k}
  \frac{d}{d\zeta} \psi_+ -\frac{\nu+{1\over 2}}{\zeta}\psi_+
- \kappa^2 \zeta \psi_+ &=&
\mathcal{M} \psi_-.
\end{eqnarray}
with eigenfunctions
\begin{eqnarray} \nonumber
\psi_+(\zeta) &\sim& z^{\frac{1}{2} + \nu} e^{-\kappa^2 \zeta^2/2}
  L_n^\nu(\kappa^2 \zeta^2) ,\\
\psi_-(\zeta) &\sim&  z^{\frac{3}{2} + \nu} e^{-\kappa^2 \zeta^2/2}
 L_n^{\nu+1}(\kappa^2 \zeta^2),
\end{eqnarray}
and  eigenvalues
\begin{equation}
\mathcal{M}^2 = 4 \kappa^2 (n + \nu + 1) .
\end{equation}

\begin{center}
\includegraphics[angle=0,width=8.0cm]{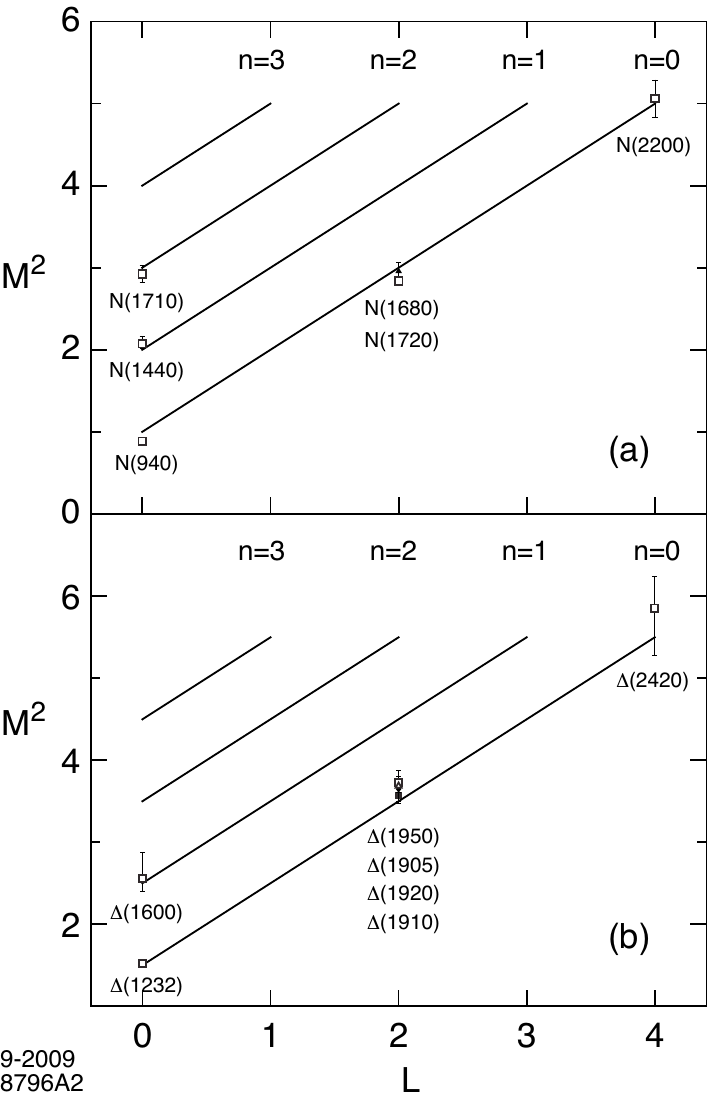}
\figcaption{\label{Baryons}{{\bf 56} Regge trajectories for  the  $N$ and $\Delta$ baryon families for $\kappa= 0.5$ GeV}
}
\end{center}

The baryon interpolating operator
$ \mathcal{O}_{3 + L} =  \psi D_{\{\ell_1} \dots
 D_{\ell_q } \psi D_{\ell_{q+1}} \dots
 D_{\ell_m\}} \psi$,  $L = \sum_{i=1}^m \ell_i$ is a twist 3,  dimension $9/2 + L$ with scaling behavior given by its
 twist-dimension $3 + L$. We thus require $\nu = L+1$ to match the short distance scaling behavior. Higher spin fermionic modes are obtained by shifting dimensions for the fields as in the bosonic case.
Thus, as in the meson sector,  the increase  in the
mass squared for  higher baryonic state is
$\Delta n = 4 \kappa^2$, $\Delta L = 4 \kappa^2$ and $\Delta S = 2 \kappa^2,$
relative to the lowest ground state,  the proton.

The predictions for the $\bf 56$-plet of light baryons under the $SU(6)$  flavor group are shown in Fig. \ref{Baryons}.
As for the predictions for mesons in Fig. \ref{VM}, only confirmed PDG~\cite{Amsler:2008xx} states are shown.
The Roper state $N(1440)$ and the $N(1710)$ are well accounted for in this model as the first  and second radial
states. Likewise the $\Delta(1660)$ corresponds to the first radial state of the $\Delta$ family. The model is  successful in explaining the important parity degeneracy observed in the light baryon spectrum, such as the $L\! =\!2$, $N(1680)\!-\!N(1720)$ degenerate pair and the $L=2$, $\Delta(1905), \Delta(1910), \Delta(1920), \Delta(1950)$ states which are degenerate
within error bars.~\cite{note1} Parity degeneracy of baryons is also a property of the hard wall model, but radial states are not well described in this model.~\cite{deTeramond:2005su}

\begin{center}
\includegraphics[angle=0,width=8.0cm]{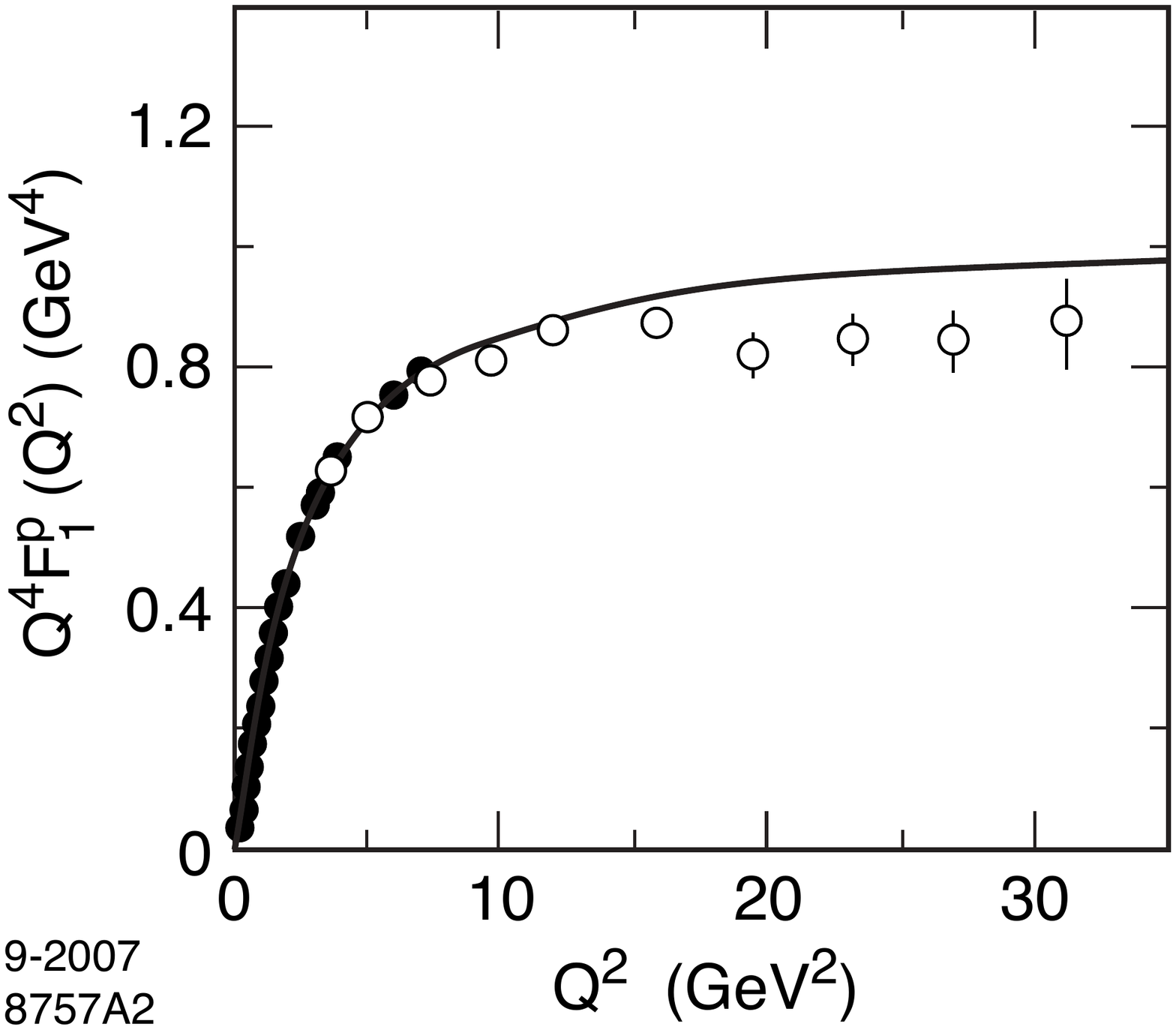}
\includegraphics[angle=0,width=7.8cm]{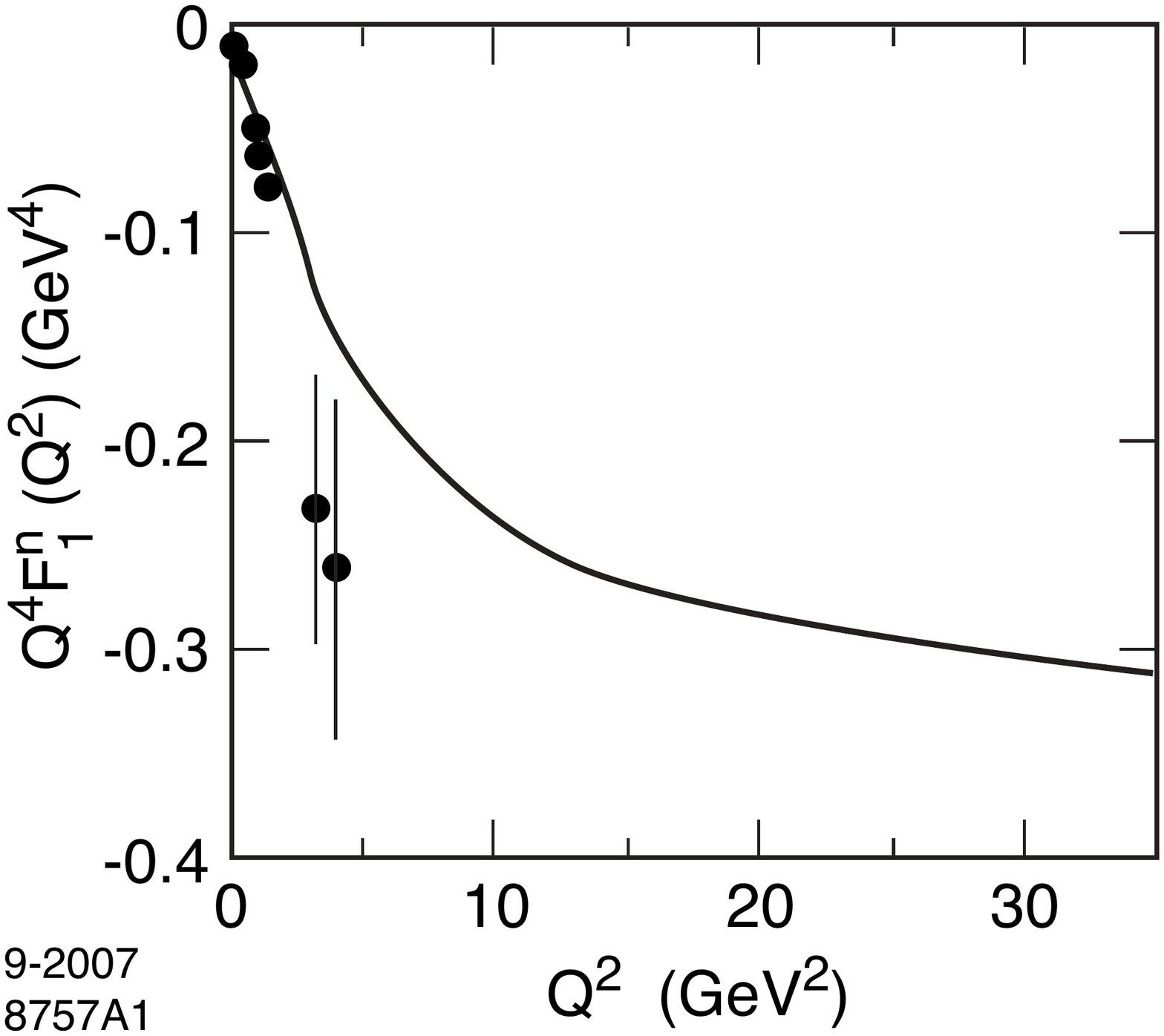}
\figcaption{
Predictions for $Q^4 F_1^p(Q^2)$ and $Q^4 F_1^n(Q^2)$ in the
soft wall model for $\kappa =  0.424$ GeV. }
 \label{fig:nucleonFF}
\end{center}

As an example  of the scaling behavior of a twist $\tau = 3$ hadron, we compute the spin non-flip
nucleon form factor in the soft wall model.~\cite{Brodsky:2008pg} The proton and neutron Dirac
form factors are given by
\begin{equation}
F_1^p(Q^2) =  \! \int  d \zeta \, J(Q, \zeta) \,
  \vert \psi_+(\zeta)\vert^2 ,
\end{equation}
\begin{equation}
F_1^n(Q^2) =  - \frac{1}{3}  \! \int  d \zeta  \,  J(Q, \zeta)
 \left[\vert \psi_+(\zeta)\vert^2 - \vert\psi_-(\zeta)\vert^2\right],
 \end{equation}
where $F_1^p(0) = 1$,~ $F_1^n(0) = 0$. The non-normalizable mode
 $J(Q,z)$ is the solution of the
AdS wave equation for the external electromagnetic current in presence of a dilaton
background and the $\exp(\pm \kappa^2 z^2)$.~\cite{Brodsky:2007hb, Grigoryan:2007my}
Plus and minus components of the twist 3 nucleon LFWF are
\begin{equation} \label{eq:PhipiSW}
\psi_+(\zeta) \!=\! \sqrt{2} \kappa^2 \, \zeta^{3/2}  e^{-\kappa^2 \zeta^2/2},  ~
\Psi_-(\zeta) \!=\!  \kappa^3 \, \zeta^{5/2}  e^{-\kappa^2 \zeta^2/2}.
\end{equation}
The results for $Q^4 F_1^p(Q^2)$ and $Q^4 F_1^n(Q^2)$   and are shown in
Fig. \ref{fig:nucleonFF}.~\cite{note3}

\section{Novel Applications of Light-Front Holography}

\begin{itemize}

\item The AdS/QCD  model
does not account for particle
creation and absorption, and thus it will break down at short distances
where hard gluon exchange and quantum corrections become important.
This semi-classical first approximation to QCD can be systematically improved  by using the complete orthonormal solutions of Eq. \ref{eq:QCDLFWE} to diagonalize the QCD light-front Hamiltonian~\cite{Vary:2009gt}  or by applying the Lippmann-Schwinger method to systematically include the QCD interaction terms, together with a variational principle. In either case, the result is the full Fock state structure of the hadron eigensolution.

\item One can model
heavy-light
and heavy hadrons by including non-zero quark masses in the LF kinetic energy
$\sum_i ({k^2_{\perp  i}+ m^2_i)/x_i}$
 as well as the effects of the one-gluon exchange potential.

\item Given
the LFWFs one can compute jet hadronization at the amplitude level from first principles.~\cite{Brodsky:2008tk}.  LF quantization also provides a distinction between static  (the absolute square of LFWFs) structure functions versus non-universal dynamic structure functions,  such as the Sivers single-spin correlation and distributions measured in diffractive deep inelastic scattering which involve final state interactions.  The origin of nuclear shadowing and process independent anti-shadowing also becomes explicit.

\item  Light-front holographic mapping of  effective classical gravity in AdS space, modified by a positive-sign dilaton background $\exp{(+ \kappa^2 z^2)}$, can be used to identify~\cite{BDT} a
 non-perturbative effective coupling $\alpha_s^{AdS}(Q^2)$ and $\beta$-functions which are in agreement with available data
 extracted from different observables and with the predictions of a class of  models with built-in confinement and lattice simulations.  The AdS/QCD  $\beta$-function shows  the transition from  perturbative to non-perturbative conformal regimes  at a momentum scale $Q \sim 1$ GeV, thus giving further support to the application of the gauge/gravity duality to the confining dynamics of strongly coupled QCD.
There are many phenomenological applications where detailed knowledge of the QCD coupling and the renormalized gluon propagator at relatively soft momentum transfer is essential.
This includes the rescattering (final-state and initial-state interactions) which
create the leading-twist Sivers single-spin correlations in
semi-inclusive deep inelastic scattering,~\cite{Brodsky:2002cx, Collins:2002kn} the Boer-Mulders functions which lead to anomalous  $\cos 2 \phi$  contributions to the lepton pair angular distribution in the unpolarized Drell-Yan reaction~\cite{Boer:2002ju} and the Sommerfeld-Sakharov-Schwinger correction to heavy quark production at threshold.~\cite{Brodsky:1995ds}
The confining AdS/QCD coupling can thus lead to a
quantitative understanding of this factorization-breaking physics.~\cite{Collins:2007nk}

\item AdS/QCD  provides  a description of chiral symmetry breaking by
using the propagation of a scalar field $X(z)$
to represent the dynamical running quark mass. The AdS
solution has the form~\cite{Erlich:2005qh,DaRold:2005zs} $X(z) = a_1 z+ a_2 z^3$, where $a_1$ is
proportional to the current-quark mass. The coefficient $a_2$ scales as
$\Lambda^3_{QCD}$ and is the analog of $\langle \bar q q \rangle$;  however,
since the quark is a color nonsinglet, the propagation of $X(z),$ and thus the
domain of the quark condensate, is limited to the region of color confinement.
Furthermore the effect of the $a_2$ term
varies within the hadron, as characteristic of an ``in-hadron" condensate.
A similar solution is found in the soft wall model in presence of a positive sign dilaton.~\cite{Zuo:2009dz}
A new perspective on the nature of quark and gluon condensates in
quantum chromodynamics is thus obtained:~\cite{Brodsky:2008be, Brodsky:2008xm, Brodsky:2008xu}  the spatial support of QCD condensates
is restricted to the interior of hadrons, since they arise due to the
interactions of confined quarks and gluons.  In  LF theory, the condensate physics is replaced by the dynamics of higher non-valence Fock states as shown by Casher and Susskind.~\cite{Casher:1974xd}  In particular, chiral symmetry is broken in a limited domain of size $1/ m_\pi$,  in analogy to the limited physical extent of superconductor phases.  This novel description  of chiral symmetry breaking  in terms of  in-hadron condensates has also been observed in Bethe-Salpeter studies.~\cite{Maris:1997hd,Maris:1997tm}
This picture also explains the
results of recent studies~\cite{Ioffe:2002be,Davier:2007ym,Davier:2008sk} which find no significant signal for the vacuum gluon
condensate. The $45$ orders of magnitude conflict of QCD with the observed value of the cosmological condensate can thus be explained.~\cite{Brodsky:2008xu}

\end{itemize}

\section{Conclusions}

We have derived a connection between a semiclassical first approximation to QCD, quantized on the light-front,
and hadronic modes propagating on a fixed AdS background. This
leads to an effective relativistic Schr\"odinger-like equation in the AdS fifth dimension coordinate $z$ (\ref{eq:QCDLFWE}).
We have show how this AdS wave equation can be derived in physical space-time as an effective equation for valence quarks in light-front quantized theory, where one identifies the AdS fifth dimension coordinate $z$ with the LF coordinate $\zeta$.

We originally derived the light-front holographic  correspondence from the identity of electromagnetic and gravitational form factors computed in AdS and LF theory.~\cite{Brodsky:2006uqa,Brodsky:2007hb,Brodsky:2008pf} Our derivation shows that the
fifth-dimensional mass $\mu$ in the AdS equation of motion
is directly related to orbital angular momentum $L$ in physical space-time. The result is physically compelling and phenomenologically successful.
The resulting Schr\"odinger-like light-front  AdS/QCD equation provides successful predictions for the light-quark meson and baryon spectra as function of hadron spin, quark angular momentum, and radial quantum number . The pion is massless for zero mass quarks in agreement with chiral invariance arguments.
The predictions for form factors are also successful. The predicted power law fall-off of hard exclusive amplitudes agrees with dimensional counting rules as required by conformal invariance at small $z$.~\cite{Brodsky:2007hb,Brodsky:2008pg}
The AdS/QCD holographic model model can be systematically improved  by using its complete orthonormal solutions to diagonalize the full QCD light-front Hamiltonian or by applying the Lippmann-Schwinger method in order to systematically include the QCD interaction terms.
We have also outlined many new applications of this formalism.

\acknowledgments{ Presented by SJB at the Fifth International Conference On Quarks and Nuclear Physics (QNP09), 21-26 Sep 2009, Beijing, China. We thank Carl Carlson,  Alexandre Deur, Josh Erlich, Stan Glazek, Hans-Guenter Dosch, Paul Hoyer,  Dae Sung Hwang, Eberhard Klempt, Mariana Kirchbach, Craig Roberts, Ivan Schmidt, Robert Shrock,  James Vary, and Fen Zuo  for helpful conversations and collaborations.
}

\end{multicols}

\vspace{-2mm}
\centerline{\rule{80mm}{0.1pt}}
\vspace{2mm}

\begin{multicols}{2}

\end{multicols}

\vspace{5mm}

\end{document}